\documentclass[aps,twocolumn,superscriptaddress,reprint]{revtex4-1}
\usepackage{graphicx,amsmath}
\usepackage{subfigure}
\usepackage{epsfig}
\usepackage{url}

\graphicspath{{figures/}}

\begin{document}

\title{Frequency Comb Generation in 300~nm Thick SiN Concentric-Racetrack-Resonators: Overcoming the Material Dispersion Limit}

\author{Sangsik Kim}
\affiliation{School of Electrical and Computer Engineering, Purdue University, West Lafayette, IN 47907 USA}
\affiliation{Birck Nanotechnology Center, Purdue University, West Lafayette, IN 47907 USA}
\affiliation{Purdue Quantum Center, Purdue University, West Lafayette, IN 47907, USA}

\author{Kyunghun Han}
\affiliation{School of Electrical and Computer Engineering, Purdue University, West Lafayette, IN 47907 USA}
\affiliation{Birck Nanotechnology Center, Purdue University, West Lafayette, IN 47907 USA}

\author{Cong Wang}
\affiliation{School of Electrical and Computer Engineering, Purdue University, West Lafayette, IN 47907 USA}

\author{Jose A. Jaramillo-Villegas}
\affiliation{School of Electrical and Computer Engineering, Purdue University, West Lafayette, IN 47907 USA}
\affiliation{Purdue Quantum Center, Purdue University, West Lafayette, IN 47907, USA}

\author{Xiaoxiao Xue}
\affiliation{School of Electrical and Computer Engineering, Purdue University, West Lafayette, IN 47907 USA}
\affiliation{Current address: Department of Electronic Engineering, Tsinghua University, Beijing 100084, China}

\author{Chengying Bao}
\affiliation{School of Electrical and Computer Engineering, Purdue University, West Lafayette, IN 47907 USA}

\author{Yi Xuan}
\affiliation{School of Electrical and Computer Engineering, Purdue University, West Lafayette, IN 47907 USA}
\affiliation{Birck Nanotechnology Center, Purdue University, West Lafayette, IN 47907 USA}

\author{Daniel E. Leaird}
\affiliation{School of Electrical and Computer Engineering, Purdue University, West Lafayette, IN 47907 USA}

\author{Andrew M. Weiner}
\affiliation{School of Electrical and Computer Engineering, Purdue University, West Lafayette, IN 47907 USA}
\affiliation{Birck Nanotechnology Center, Purdue University, West Lafayette, IN 47907 USA}
\affiliation{Purdue Quantum Center, Purdue University, West Lafayette, IN 47907, USA}

\author{Minghao Qi}
\email[]{mqi@purdue.edu}
\affiliation{School of Electrical and Computer Engineering, Purdue University, West Lafayette, IN 47907 USA}
\affiliation{Birck Nanotechnology Center, Purdue University, West Lafayette, IN 47907 USA}
\affiliation{Purdue Quantum Center, Purdue University, West Lafayette, IN 47907, USA}

\begin{abstract}
Kerr nonlinearity based frequency combs
and solitons
have been generated from on-chip optical microresonators with high quality factors and global or local anomalous dispersion.
However, fabrication of such resonators usually requires materials and/or processes that are not standard in semiconductor manufacturing facilities.
Moreover, in certain frequency regimes such as visible and ultra-violet,
the large normal material dispersion makes it extremely difficult to achieve anomalous dispersion.
Here we present a concentric racetrack-shaped resonator that achieves anomalous dispersion in a 300~nm thick silicon nitride film,
suitable for semiconductor manufacturing but previously thought to result only in waveguides with high normal dispersion,
a high intrinsic $Q$ of 1.5~million, and a novel mode-selective coupling scheme that allows coherent combs to be generated.
We also provide evidence suggestive of soliton-like pulse formation in the generated comb.
Our method can achieve anomalous dispersion over moderately broad bandwidth for resonators at almost any wavelength
while still maintaining material and process compatibility with high-volume semiconductor manufacturing.
\end{abstract}

\maketitle 

High quality factor ($Q$) microresonators have been used for Kerr frequency comb (or ``microcomb'')
\cite{kippenberg2011microresonator,del2007optical,razzari2010cmos,levy2010cmos,herr2012universal,ferdous2011spectral,
liu2014investigation,xue2015mode,ramelow2014strong,xue2015normal,ramelow2014strong,xuan2016high}
and soliton \cite{herr2014temporal,wang2016intracavity,karpov2016universal,brasch2016photonic,kordts2016higher}
generations.
To initiate the combs, anomalous dispersion is required for the modulation instability \cite{matsko2005optical,matsko2012hard,hansson2013dynamics},
and typical approaches rely on the engineering of structural dispersions,
which in many cases cause fabrication challenges, $e.g.$, very thick films prone to cracking,
and fail in many spectrum ranges such as the visible and ultraviolet where most materials show high normal dispersion.

Nevertheless, frequency combs at a relatively low normal dispersion regime have been reported previously
\cite{liu2014investigation,xue2015mode,ramelow2014strong,xue2015normal},
and the general idea is that mode coupling between different sets of resonating modes may cause strong but highly localized anomalous dispersion,
allowing modulation instability.
Such mode couplings have been observed between higher order modes \cite{liu2014investigation,xue2015mode} or different polarizations \cite{ramelow2014strong},
yet these are more phenomenal observations and rely on accidental mode coupling.
Dual-coupled resonators with thermal tuning also have been used to control the mode interaction \cite{xue2015normal},
but, the induced anomalous dispersion is still highly localized.
Furthermore, in the Si$_3$N$_4$ platform, which allows potential CMOS integration,
these schemes still require thick Si$_3$N$_4$, which leads to significant film stress.
Current methods to mitigate the stress rely on either deep trenches
which will result in difficulties in subsequent processes such as polishing and resist spinning \cite{nam2012patterning,luke2013overcoming},
or the ``photonic damascene'' process which may cause issues in feature size and process control \cite{pfeiffer2016photonic}.
The use of compound-ring resonators to induce anomalous dispersion has been proposed with numerical simulation \cite{soltani2015enabling}.
However, the proposed scheme still requires thick Si$_3$N$_4$ film and
is in general very sensitive to fabrication imperfections, making realization difficult.
Furthermore, achieving high $Q$s and efficient pump coupling for the comb-generating mode is also critical for realizing frequency comb generation.

In this work, we present a concentric-racetrack-resonator design
that can engineer the resonator dispersion with a high degree of control,
and experimentally demonstrate strong anomalous dispersion and then frequency comb and soliton generation.
To prove the concept of overcoming the material dispersion limit,
we chose a thin (300~nm) Si$_3$N$_4$ platform, which shows high normal dispersion at near infrared.
An adiabatically tapered concentric-racetrack resonator allows for selective excitation of the mode
that shows anomalous dispersion, and coherent Kerr frequency combs are demonstrated.
Furthermore, using two free-running lasers as pumps, we achieved a broad and smooth comb state with a Cherenkov radiation peak and a single soliton-like pulse.

\begin{figure*}[!ht]
\begin{center}
{\includegraphics[width=0.9\textwidth]{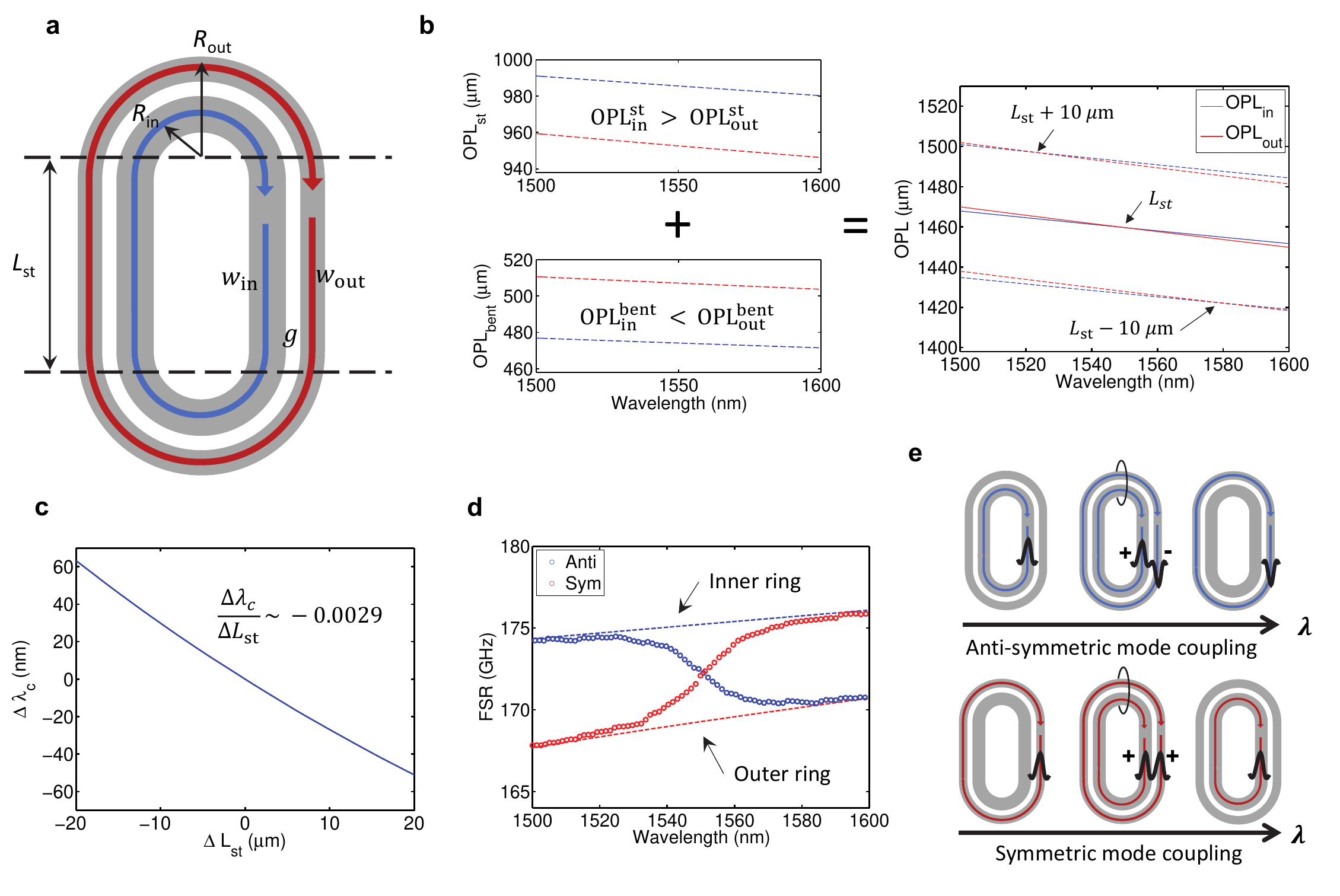}}
\caption{
{\bf Optical path length engineering with a concentric-racetrack-resonator.}
{\bf a,} Schematic of a concentric-racetrack-resonator.
The device height $h$ is set to be 300~nm,
and other geometric parameters are $R_{\rm out}=50~\mu$m, $g=900$~nm, $w_{\rm in}=2750$~nm, $w_{\rm out}=1200$~nm, and $L_{\rm st}=300~\mu$m.
{\bf b,} Optical path length engineering of inner (OPL$_{\rm in}$, blue) and outer (OPL$_{\rm out}$, red) resonant modes.
{\bf c,} Shift in OPL matching points with different $L_{\rm st}$.
{\bf d,} Calculated FSRs of symmetric (red circles) and anti-symmetric (blue circles) resonant modes.
{\bf e,} Illustrations of symmetric (upper) and anti-symmetric (lower) mode couplings.
}
\label{fig1}
\end{center}
\vspace{-3mm}
\end{figure*}
Figure~\ref{fig1}a gives the schematic of a racetrack-type concentric-resonator showing two fundamental TE modes, TE$_0^{\rm in}$ (blue) and TE$_0^{\rm out}$ (red).
With the geometric parameters listed in Fig.~\ref{fig1} caption,
the mode overlap is negligible in both the straight and curved sections,
and in the absence of resonant coupling, the modes reside mainly in the inner and outer racetracks, respectively.
However, when the round-trip optical path lengths OPL$_{\rm in}$ and OPL$_{\rm out}$ of the inner and outer racetracks are matched (OPL$_{\rm in}=$OPL$_{\rm out}$),
or in other words, the phase matching condition is met,
there is nontrivial crosstalk between inner and outer modes exchanging their optical paths, and this process leads to resonant mode-coupling.
Figure~\ref{fig1}b shows the detailed OPL matching between inner and outer resonators with OPLs at bent (OPL$^{\rm bent}$) and straight (OPL$^{\rm st}$) sections.
The OPL$_{\rm in}$ and OPL$_{\rm out}$ can be represented as the following,
\begin{eqnarray*}
{\rm OPL_{in}}&=&{\rm OPL_{in}^{st}}+{\rm OPL_{in}^{bent}} =2 L_{\rm st}{n_{\rm in}^{\rm st}}+2\pi R_{\rm in}{n_{\rm in}^{\rm bent}}\\
{\rm OPL_{out}}&=&{\rm OPL_{out}^{st}}+{\rm OPL_{out}^{bent}} =2 L_{\rm st}{n_{\rm out}^{\rm st}}+2\pi R_{\rm out}{n_{\rm out}^{\rm bent}}
\label{eq2}
\end{eqnarray*}
where $L_{\rm st}$ is the length of the straight section and $R$ is the effective bending radius of each mode.
The $n$ is the effective refractive index of each waveguide mode, and there are four different $n$:
inner and outer rings at both the straight and bent areas.
In general, at the bending area, ${\rm OPL_{out}^{bent}}$ is greater than ${\rm OPL_{in}^{bent}}$ due to the larger bending radius,
while at the straight section, ${\rm OPL_{in}^{st}}$ is greater than ${\rm OPL_{out}^{st}}$ due to the larger $n$ with a larger waveguide width.
This opposite OPL difference ($\Delta$OPL) allows us to match the round-trip OPLs at desired wavelengths almost independently of the coupling strength.
The OPL matching point (wavelength) can be engineered by changing the $L_{\rm st}$,
and Fig~\ref{fig1}c shows the shift in OPL matching wavelength, $\Delta\lambda_c / \Delta L_{\rm st}\sim-0.0029$.
This independent and very fine control of the mode coupling point distinguishes our method from the scheme proposed in \cite{soltani2015enabling},
where phase matching between two concentric rings, $i.e.$, without the straight sections, was to be achieved by varying the waveguide widths,
which requires precise width control at sub-10~nm precision, much more difficult than achieving the right $L_{\rm st}$.

As a result of the OPL exchange between the inner and outer rings,
the resonant mode-coupling generates symmetric ($\omega_{\rm s}$) and anti-symmetric ($\omega_{\rm a}$) hybridized modes,
the resonant frequencies of which can be represented by
\begin{equation}
\omega_{\rm s,a}=\frac{\omega_{\rm m}+\omega_{\rm n}}{2} \pm \sqrt{\frac{(\omega_{\rm m}-\omega_{\rm n})^2}{4}+\kappa^2_\omega},
\label{eq1}
\end{equation}
where $\kappa_{\omega}$ is the coupling coefficient 
and $\omega_{\rm m}$ and $\omega_{\rm n}$ are the resonant modes at the inner and outer racetracks without mode coupling, which can be found by
${\rm OPL_{\rm in}}={\rm m}\lambda_{\rm m}$ and ${\rm OPL_{\rm out}}={\rm n}\lambda_{\rm n}$ (m,n: real integers).
The calculated free spectral ranges (FSRs) of each mode are plotted in Fig.~\ref{fig1}d,
where the FSR of the anti-symmetric mode (blue circles) decreases as the wavelength increases.
This indicates anomalous dispersion in the anti-symmetric mode ($D_\lambda=\frac{\partial}{\partial\lambda}(\frac{1}{\rm FSR_\nu(\lambda)L})>0$).
Note that the dispersion of the inner and outer racetrack modes in spectral regions away from the resonant mode coupling,
which are asymptotic to the dashed lines, are both normal.
The anomalous dispersion comes only from the resonant mode coupling.
An intuitive understanding of these FSR curves is that the anti-symmetric mode evolves from the inner racetrack to the outer one as the wavelength increases,
while it's opposite for the symmetric mode (Fig.~\ref{fig1}e).
We note that the exact dispersion at each segment of the resonator is not critical
as long as there is an overall balance of the OPL of the two modes (or the accumulated round-trip phase delays are matched).

\begin{figure*}[!ht]
\begin{center}
{\includegraphics[width=1.0\textwidth]{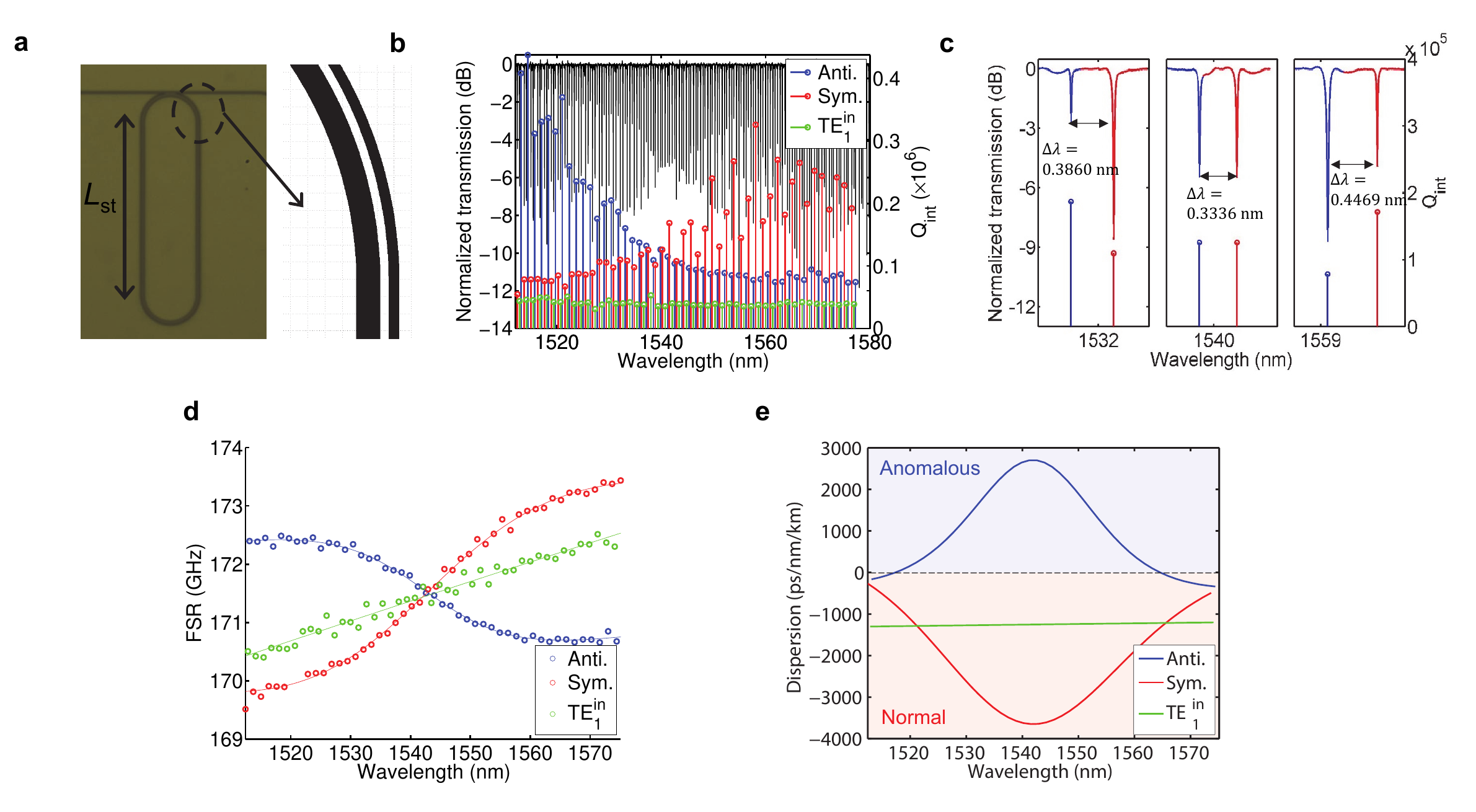}}
\caption{
{\bf Resonant mode-coupling in a concentric-racetrack-resonator.}
{\bf a,} Optical image and zoomed-in layout of the concentric-racetrack resonator. The nominal geometric parameters are the same as in Fig.~\ref{fig1}a.
{\bf b,} Normalized transmission spectrum and fitted intrinsic quality factors of each mode:
anti-symmetric (blue), symmetric (red), and 1$^{\rm st}$ order TE mode at inner ring TE$^{\rm in}_1$ (green).
{\bf c,} Zoomed-in view of {\bf (c)} before, at, and after the mode coupling point.
{\bf d, e,} Characterized {\bf (d)} free spectral range and {\bf (e)} dispersion of each mode.
}
\label{fig2}
\end{center}
\vspace{-3mm}
\end{figure*}
Figure~\ref{fig2}a shows the optical image of the fabricated device and
Fig.~\ref{fig2}b shows the normalized transmission (black) and characterized intrinsic quality factors ($Q_{\rm int}$), which correspond to the round-trip propagation losses.
There are three families of resonant modes; we can assign them as
anti-symmetric (blue), symmetric (red), and TE$^{\rm in}_1$ (green) based on the difference between their trends of $Q_{\rm int}$ vs wavelength.
First, the mode that has relatively low and constant $Q_{\rm int}$ is assigned to TE$^{\rm in}_1$
(note that the $w_{\rm out}=1200$~nm is too narrow to support TE$_1$ mode),
which means that the round trip loss of TE$^{\rm in}_1$ mode does not change appreciably with the wavelength,
a reasonable assumption when the mode profile and optical path do not change drastically within the wavelength range.
The round-trip losses, or $Q_{\rm int}$, of the other two modes, however, change significantly in the measured wavelength range.
According to Fig.~\ref{fig1}e, the anti-symmetric mode at shorter wavelengths is well confined in the inner racetrack, which is much wider than the outer one.
In general, a wider waveguide offers higher field confinement and yields higher $Q_{\rm int}$ due to the reduction of scattering losses from the sidewall roughness.
Therefore, the $Q_{\rm int}$ is larger at shorter wavelengths and decreases as the wavelength increases
since the anti-symmetric mode (shown in blue in Fig.~\ref{fig2}b) evolves to the outer racetrack, which has a smaller width.
Meanwhile, the opposite trend is observed for the symmetric mode (shown in red in Fig.~\ref{fig2}b).
Figure~\ref{fig2}c is the zoomed-in view of a few resonant modes in Fig.~\ref{fig2}b
and shows that the extinction ratios of the resonances also exhibit opposite trends vs wavelength.
For the anti-symmetric mode, the on-resonance extinction increases with the wavelength,
suggesting an increase of coupling between the bus waveguide and the resonator in the under coupled regime.
This is consistent with our determination that the anti-symmetric mode evolves from the inner racetrack to the outer one.
Again, the opposite trend is observed for the symmetric mode.

After unequivocally determining the mode families, we characterized the FSRs and dispersions of each mode in Figs~\ref{fig2}d and \ref{fig2}e, respectively.
The TE$^{\rm in}_1$ (green) mode has high normal dispersion (around -1200 ps/nm/km),
which is a typical dispersion value for 300~nm thick Si$_3$N$_4$ resonators that prevents frequency comb generation.
The FSRs for the symmetric (red) and anti-symmetric (blue) modes clearly show different trends,
with the desirable anomalous dispersion occurring for the anti-symmetric mode and normal dispersion for the symmetric mode.
The measured FSRs are qualitatively consistent with the simulated ones shown in Fig.~\ref{fig1}d;
in particular the resonant mode-coupling point, or the wavelength at which OPL$_{\rm in}=$OPL$_{\rm out}$, matches fairly well.
This is important as the anomalous dispersion band should reside within the gain bandwidth of the erbium-doped fibre amplifiers (EDFAs)
to have access to strong optical pumping.

\begin{figure*}[!ht]
\begin{center}
{\includegraphics[width=1.0\textwidth]{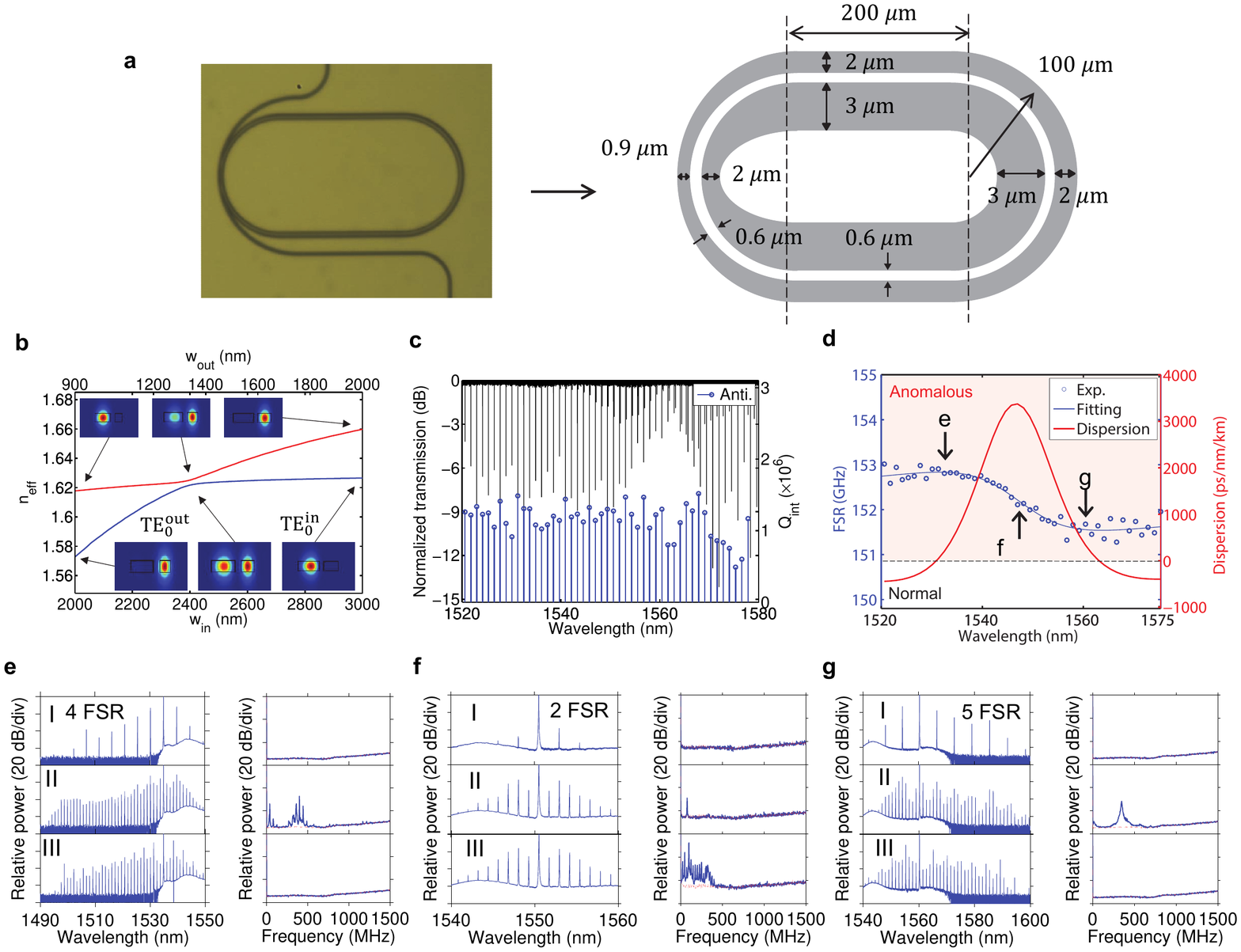}}
\caption{
{\bf Frequency comb generations with a concentric-racetrack-resonator.}
{\bf a,} Optical image and zoomed-in layout (not to scale) of the concentric-racetrack-resonator.
The film thickness $h$ remains 300~nm, $R_{\rm out}=100~\mu$m, $g=600$~nm, $L_{\rm st}=200~\mu$m, $w_{\rm in}=3000$~nm and $w_{\rm out}=2000$~nm
(except in the tapered portion of the device).
The curved section on the left side of the resonator is tapered so that the widths of the inner and outer rings change adiabatically from
$w_{\rm in}=3000$ to 2000 and then back to 3000~nm, and $w_{\rm out}=2000$ to 900 and then back to 2000~nm, respectively.
The gap between the inner and outer bends remain constant at $g=600$~nm.
{\bf b,} Simulated effective refractive indices through the tapered section.
{\bf c,} Normalized transmission spectrum and fitted intrinsic quality factors of the anti-symmetric mode (blue).
{\bf d,} Free spectral range (blue) and corresponding dispersion (red) of the anti-symmetric mode.
{\bf e, f, g,} Comb spectra and RF noises with pumping at different resonant modes of
{\bf (e)} $\lambda_{\rm p}\sim1535.0$~nm, {\bf (f)} $\lambda_{\rm p}\sim1550.6$~nm, and {\bf (g)} $\lambda_{\rm p}\sim1560.3$~nm, respectively.
From I to III, the pump laser is tuned into resonance by red-shifting the pump wavelengths.
The pump powers are {\bf (e)} 425~mW, {\bf (f)} 495~mW, and {\bf (g)} 485~mW, at the bus waveguide.
The red lines on the RF measurements are the background noise of the electronic spectrum analyzer (ESA).
}
\label{fig3}
\end{center}
\vspace{-3mm}
\end{figure*}
For microcomb generation, in addition to the dispersion requirements,
a higher $Q_{\rm int}$ ($>10^6$) to build up optical power within the cavity is also necessary.
We increased the bending radius and racetrack waveguide widths to reduce the scattering losses from the sidewall roughness.
Using a fabrication method we recently developed \cite{xuan2016high},
we were able to achieve high intrinsic $Q$s in the 1-2 million range, and we observed frequency comb generation (see Fig.~S1 in Supplementary Information).
Nevertheless, the microcombs generated are not evenly spaced, and they have high Radio-frequency (RF) beating noise characteristic of type~II combs
\cite{herr2012universal,ferdous2011spectral}.
The main reason is that our resonator is inherently multi-mode, and increasing the racetrack width favors the excitation of the symmetric or other high-order modes.
This leads to complicated comb spectrum and prevents the mode-locking.

Fortunately, as we mentioned before, our phase-matching condition is highly tolerant
and requires only that there be an overall balance of optical path length for the two coupling resonant modes.
Therefore, we can adiabatically taper the region of the resonator that couples to the bus waveguide to excite the anti-symmetric mode selectively
(Fig.~\ref{fig3}a, also see Methods for detailed geometry) within certain wavelength range.
According to Fig.~\ref{fig1}e, in order to evolve into the antisymmetric mode,
the mode should be concentrated in the inner racetrack (TE$^{\rm in}_0$ like) at wavelengths shorter than the mode coupling point.
However, due to the proximity of the outer racetrack to the bus waveguide, power will always be coupled to the outer racetrack with much higher efficiency. Therefore, the taper should transform the mode from the outer racetrack (TE$^{\rm out}_0$ like) to the inner one (TE$^{\rm in}_0$ like).
Figure~\ref{fig3}b shows the simulated effective refractive indices of the designed adiabatically tapered concentric racetrack bends (at $\lambda_0=1550$~nm).
Insets show the mode profiles at three different pairs of $w_{\rm in}$ and $w_{\rm out}$.
With this tapering, the pump light is first coupled to the TE$^{\rm out}_0$ mode due to the proximity of the outer racetrack to the bus waveguide.
Then the TE$^{\rm out}_0$ mode transforms into the TE$^{\rm in}_0$ mode as light propagates through the tapered section,
following the blue line in Fig.~\ref{fig3}b.
Our tapering design also provides a mode filtering function to prevent the excitation of
other unnecessary higher order and symmetric modes and improves the coherence of the comb \cite{kordts2016higher}.
Figure~\ref{fig3}c shows the transmission spectrum and $Q_{\rm int}$ of the antisymmetric mode.
As expected, compared to Fig.~\ref{fig2}b, the spectrum shows almost a single resonant mode
at wavelengths shorter than the OPL matching wavelength ($\sim1548$~nm).
Figure~\ref{fig3}d shows the characterized FSR (blue) and dispersion (red) of the dominant resonant mode.
Notice the anomalous dispersion that is due to anti-symmetric coupling at the single mode regime ($1530-1548$~nm).
At longer wavelengths, two mode families exist and we chose the resonance dips that form a smooth dispersion spectrum (Fig.~\ref{fig3}d, red curve)
when combined with the dispersion derived from the single mode regime ($<1548$~nm).
This suggests that our resonator design allows the excitation of the anti-symmetric mode at wavelengths longer than the OPL matching point
(supported by comb generation shown in Fig.~\ref{fig3}g),
even though the tapers favor the symmetric mode which manifests itself as the deep resonance dips in the wavelength range of $1566-1580$~nm.
The average $Q_{\rm int}$ does not show trends similar to Fig.~\ref{fig1}b,
suggesting that the $Q_{\rm int}$ is limited by the introduction of the tapers in the coupling region.
Despite that, a high $Q_{\rm int}$ around $1\times10^6$ is achieved across the measurement range.

To generate the combs, we used the conventional method of laser tuning that slowly scans from blue to red.
We pumped at three different resonances:
at wavelengths shorter than ($1535.0$~nm), at ($1550.6$~nm), and longer than ($1560.4$~nm) the resonant mode coupling,
and Figs~\ref{fig3}e, \ref{fig3}f, and \ref{fig3}g show the corresponding comb spectra, respectively.
Low frequency RF noise spectra are also plotted.
The comb lines expand to a shorter wavelength when we pump at $1535.0$~nm (Fig.~\ref{fig3}e),
while they expand to a longer wavelength when we pump at $1560.4$~nm (Fig.~\ref{fig3}g).
In both these cases pumping occurs at wavelengths with close to zero dispersion, and the comb expands asymmetrically,
extending predominantly into the region with small and approximately wavelength-independent normal dispersion.
A mode-locking transition, in which a noisy comb switches to a low noise phase-locked state \cite{liang2014generation}, was also observed in these combs.
The low RF noise in the mode-locked states suggests high coherence;
this is confirmed by line-by-line shaping and autocorrelation measurements (see Fig.~S2 in Supplementary Information) \cite{ferdous2011spectral}.
On the other hand, the microcomb in Fig.~\ref{fig3}f, where pumping occurs in the resonant mode coupling region, shows very different behavior.
Here the comb exhibits a symmetric profile with very limited bandwidth, and is restricted to the region
where the dispersion is large, anomalous, and strongly wavelength-dependent.
Furthermore, this comb exhibits high RF noise,
and the autocorrelation data (Fig.~S2) indicate poor compressibility under line-by-line shaping, a hallmark of incoherence.

\begin{figure*}[!ht]
\begin{center}
{\includegraphics[width=0.7\textwidth]{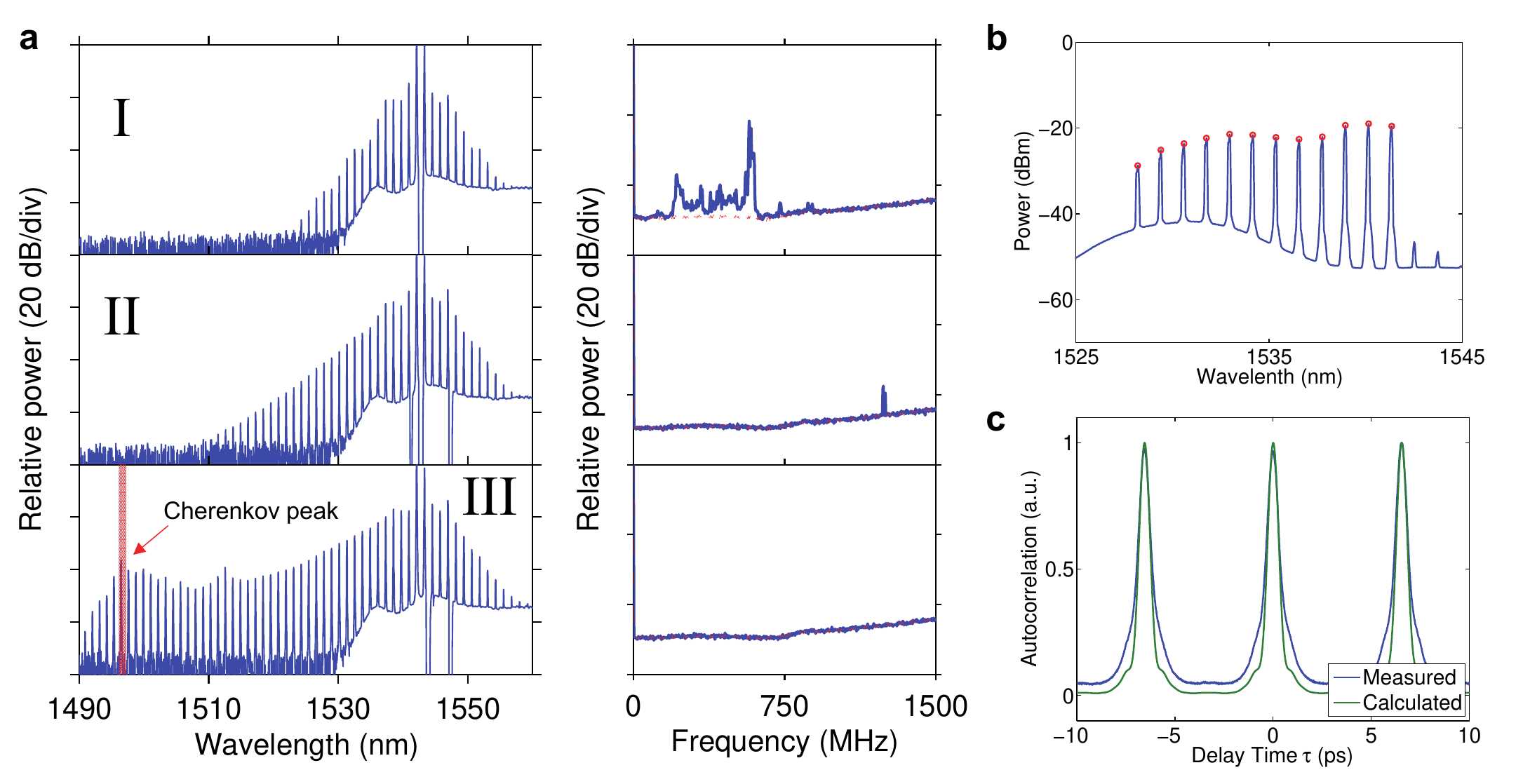}}
\caption{
{\bf Soliton generation with bichromatic pump.}
{\bf a,} Evolution of comb spectrum (left) and corresponding RF noise (right), with bichromatic pumping at two resonances separated by 1-FSR:
$\lambda_{\rm p1}\sim1542.2$~nm (scanning) and $\lambda_{\rm p2}\sim1543.4$~nm (fixed).
${\rm I}$, ${\rm II}$, and ${\rm III}$ are at slightly tuned $\lambda_{\rm p1}$ while scanning from blue to red.
{\bf b,} Spectrum after pulse shaper and EDFA.
{\bf c,} Autocorrelation of comb from {\bf (b)} with fiber dispersion compensated via the pulse shaper (blue) and autocorrelation computed assuming experimental spectrum with flat spectral phase (green).
}
\label{fig4}
\end{center}
\vspace{-3mm}
\end{figure*}

%
It has recently been demonstrated that the initially noisy,
Type II combs generated from resonators with global anomalous dispersion can transition into a coherent,
low noise state associated with the formation of solitons \cite{herr2014temporal,wang2016intracavity,karpov2016universal}.
Here we use bichromatic pumping \cite{papp2013parametric,liu2013dual}
to achieve comb operation strongly suggestive of soliton formation in the high anomalous dispersion regime
associated with resonant mode coupling in our concentric racetrack resonator.
Our experiment utilizes two free-running pump lasers, one fixed in the high anomalous dispersion regime at $\lambda_{\rm p2}\sim1543.4$~nm,
the other scanned approximately 1 FSR away ($\lambda_{\rm p1}\sim1542.2$~nm).
Figure~\ref{fig4}a shows the evolution of the comb spectra and the corresponding RF noise measurements.
In the panel denoted III, the RF noise has dropped out, and the spectrum is very smooth and has extended significantly.
These attributes are indicative of a phase-locking transition, which we believe coincides with formation of single solitons.
As evidence, we note that the spectrum develops a peak around 1496~nm, which is in the normal dispersion regime.
We identify this as a Cherenkov peak, attributed to phase matching of a soliton
with the dispersive wave expected for frequency components in the normal dispersion regime \cite{brasch2016photonic}.
The peak position ($\omega_{\rm c}$) agrees closely with the theoretical prediction of
$\omega_{\rm c}=\omega_{\rm p2}-3\frac{\beta_2(\omega_{\rm p2})}{\beta_3(\omega_{\rm p2})}=1.2589\times10^{15}$~rad/s ($1496.3$~nm) \cite{erkintalo2012cascaded,brasch2016photonic},
expected for Cherenkov radiation.
To further investigate, we performed a time-domain autocorrelation measurement.
A pulse shaper was used to select and equalize a group of comb lines  and to compensate
for the dispersion of the fibre link (no additional phase correction was applied to the comb lines selected, see Methods).
Figure~\ref{fig4}b is the spectrum that is used in autocorrelation, and Fig.~\ref{fig4}c shows the autocorrelation measurement result (blue).
The autocorrelation trace consists of one peak per round trip with high contrast, indicative of isolated short pulses.
Furthermore, the experimental trace is well matched by the autocorrelation simulated under the assumption that the comb is bandwidth-limited,
$i.e.$, has constant spectral phase (Fig.~\ref{fig4}c, green).
These results further suggest the attainment of a soliton-like coherent state.

In summary, we present a complete set of design guidelines to simultaneously achieve
1) well-controlled phase matching of two resonating modes in a concentric-racetrack-resonator;
2) resonant mode-coupling to significantly alter the FSR of the anti-symmetrically coupled resonant mode to achieve anomalous dispersion in a relatively broad wavelength range, and 3) preferential coupling to the anti-symmetric mode and filtering out of the symmetric and other uncoupled modes.
These guidelines are robust and tolerant enough to allow us to experimentally demonstrate concentric-racetrack-resonators
with intrinsic $Q$s over $1\times10^6$ on a 300~nm thick Si$_3$N$_4$ platform which can be readily fabricated in a commercial CMOS foundry.
Despite the high normal dispersion in 300~nm thick Si$_3$N$_4$ waveguides,
we successfully generated coherent Kerr frequency combs with smooth spectral shape and a Cherenkov peak in the near infrared.
Together with autocorrelation measurement of a portion of the comb spectrum, our results suggest the achievement of a single bright Kerr soliton.
Therefore, our new resonators exhibit almost all the important phenomena of the microcombs in the anomalous dispersion regime, albeit with a smaller comb bandwidth.
Nevertheless, in many applications such as telecommunications and microwave photonics,
limiting the comb bandwidth might be preferred since it could boost the energy of individual comb lines.
Moreover, our methodology should allow microcombs to be generated in many spectral ranges that are currently not feasible due to the high normal material dispersion,
$e.g.$, visible and ultraviolet.
Finally, the critical dimensions in our design are above 500~nm, therefore high-resolution lithography is not absolutely required.
Therefore, we expect our methodology and demonstration to have considerable impact for on-chip frequency comb research
and to pave the way for potential practical applications
because the fabrication is now feasible in a commercial CMOS foundry without the latest lithography capability.

\section*{Methods}
{\bf Device fabrication}
A $<$100$>$-oriented 4-inch silicon wafer was used as a substrate, and a 3~$\mu$m-thick silicon dioxide was thermally grown in a wet oxidation furnace.
A 300~nm thick stoichiometric Si$_3$N$_4$ film was deposited by a low-pressure chemical vapor deposition (LPCVD).
The concentric-racetrack-resonator was patterned by a 100~kV electron-beam lithography tool with hydrogen silsesquioxane (HSQ) negative tone e-beam resist.
The nitride film was etched by an inductively coupled plasma reactive ion etcher (ICP-RIE) with CHF$_3$ and O$_2$.
The HSQ was removed by a buffered oxide etcher to avoid the influence of the e-beam resist.
The patterned device was annealed at $1150~^{\circ}$C for 3 hours in N$_2$ ambient gas.
A 3~$\mu$m-thick silicon dioxide top cladding was deposited as a low-temperature oxide in an LPCVD furnace.
Using photolithography, the edges of the inverse taper fiber couplers were defined and etched through the substrate for 90~$\mu$m with ICP-RIE.
The fiber edge coupler efficiency is 6~dB/facet.

{\bf Device geometry}
All the device heights $h$ are set to be 300~nm,
and other geometric parameters for the device in Figs.~\ref{fig1} and \ref{fig2} are
$h=300$~nm, $R_{\rm out}=50~\mu$m, $g=900$~nm, $w_{\rm in}=2750$~nm, $w_{\rm out}=1200$~nm, and $L_{st}=300~\mu$m.
The bus waveguide width is 1.2~$\mu$m, and the gap between bus and outer ring is 300~nm.
For the device in Fig.~\ref{fig3} the parameters are
$h=300$~nm, $R_{\rm out}=100~\mu$m, $g=600$~nm, $L_{st}=200$~nm, $w_{\rm in}=3000$~nm and $w_{\rm out}=2000$~nm (in the untapered portion of the device).
The region of the device with coupling to the bus waveguide (left side of Fig.~\ref{fig3}a) is tapered
so that the widths of the inner and outer rings change adiabatically from $w_{\rm in}=3000$ to 2000 to 3000~nm
and $w_{\rm out}=2000$ to 900 to 2000~nm, respectively.
The gap size $g$ between inner and outer rings is held constant $g=600$~nm.
The arc length over which the taper occur is 300.22~$\mu$m for inner ring and 311.96~$\mu$m for outer ring.
The bus waveguide width is 1.2~$\mu$m, the gap between bus and outer ring is 200~nm, and the bending radius of bus waveguide is 130~$\mu$m.

{\bf Line-by-line pulse shaping and time domain autocorrelation measurements}
As described in ref.~\cite{ferdous2011spectral},
we selected a few comb lines and conducted line-by-line phase correction to form a transform-limited pulse.
The compressed pulse was amplified with an EDFA to compensate for the losses from the pulse shaper and link,
then measured with an autocorrelator based on noncollinear second harmonic generation.
The agreement of the compressed signal and calculated signal with flat phases (calculated based on the intensity of the comb lines in the optical spectrum)
indicates that the pulse is close to bandwidth limited \cite{ferdous2011spectral}.
The same measurement setup was used for the autocorrelation in Fig.~\ref{fig4}c, but without any phase correction on the comb lines;
here the pulse shaper was used to compensate only for the dispersion of the fibre link and the measurement system.
For the dispersion compensation, a mode-locked fibre laser was used as a reference.
These methods of using a pulse shaper to form a bandwidth-limited pulse and to compensate for the dispersion of the link
have been widely used in previous research and verified with other methods \cite{supradeepa2009optical,del2015phase,ferdous2011spectral,xue2015mode}.

\bibliography{CRR_arxiv}

\section*{Acknowledgments}
This work was supported in part by the National Science Foundation under grant ECCS-1509578,
by the Air Force Office of Scientific Research under grant FA9550-15-1-0211,
and by the DARPA PULSE program through grant W31P40-13-1-0018 from AMRDEC.

\section*{Author contributions}
S.K. conceived the idea, performed the numerical simulations, and designed the device layouts.
S.K. led the characterization with assistance from C.W., J.A.J., and X.X.
K.H. and Y.X. fabricated the devices.
S.K., M.Q., and A.M.W. wrote the manuscript.
All discussed and commented on the results.
The project was organized and coordinated by M. Q. and A.M.W.

\end{document}


\title{Supplementary Information to ``Frequency Comb Generation in 300~nm Thick SiN Concentric-Racetrack-Resonators: Overcoming the Material Dispersion Limit''}

\author{Sangsik Kim}
\affiliation{School of Electrical and Computer Engineering, Purdue University, West Lafayette, IN 47907 USA}
\affiliation{Birck Nanotechnology Center, Purdue University, West Lafayette, IN 47907 USA}
\affiliation{Purdue Quantum Center, Purdue University, West Lafayette, IN 47907, USA}

\author{Kyunghun Han}
\affiliation{School of Electrical and Computer Engineering, Purdue University, West Lafayette, IN 47907 USA}
\affiliation{Birck Nanotechnology Center, Purdue University, West Lafayette, IN 47907 USA}

\author{Cong Wang}
\affiliation{School of Electrical and Computer Engineering, Purdue University, West Lafayette, IN 47907 USA}

\author{Jose A. Jaramillo-Villegas}
\affiliation{School of Electrical and Computer Engineering, Purdue University, West Lafayette, IN 47907 USA}
\affiliation{Purdue Quantum Center, Purdue University, West Lafayette, IN 47907, USA}

\author{Xiaoxiao Xue}
\affiliation{School of Electrical and Computer Engineering, Purdue University, West Lafayette, IN 47907 USA}
\affiliation{Current address: Department of Electronic Engineering, Tsinghua University, Beijing 100084, China}

\author{Chengying Bao}
\affiliation{School of Electrical and Computer Engineering, Purdue University, West Lafayette, IN 47907 USA}

\author{Yi Xuan}
\affiliation{School of Electrical and Computer Engineering, Purdue University, West Lafayette, IN 47907 USA}
\affiliation{Birck Nanotechnology Center, Purdue University, West Lafayette, IN 47907 USA}

\author{Daniel E. Leaird}
\affiliation{School of Electrical and Computer Engineering, Purdue University, West Lafayette, IN 47907 USA}

\author{Andrew M. Weiner}
\affiliation{School of Electrical and Computer Engineering, Purdue University, West Lafayette, IN 47907 USA}
\affiliation{Birck Nanotechnology Center, Purdue University, West Lafayette, IN 47907 USA}
\affiliation{Purdue Quantum Center, Purdue University, West Lafayette, IN 47907, USA}

\author{Minghao Qi}
\email[]{mqi@purdue.edu}
\affiliation{School of Electrical and Computer Engineering, Purdue University, West Lafayette, IN 47907 USA}
\affiliation{Birck Nanotechnology Center, Purdue University, West Lafayette, IN 47907 USA}
\affiliation{Purdue Quantum Center, Purdue University, West Lafayette, IN 47907, USA}

\maketitle 

\section{Type~II comb with a concentric-racetrack-resonator}
\begin{figure}[!ht]
\begin{center}
{\includegraphics[width=0.9\textwidth]{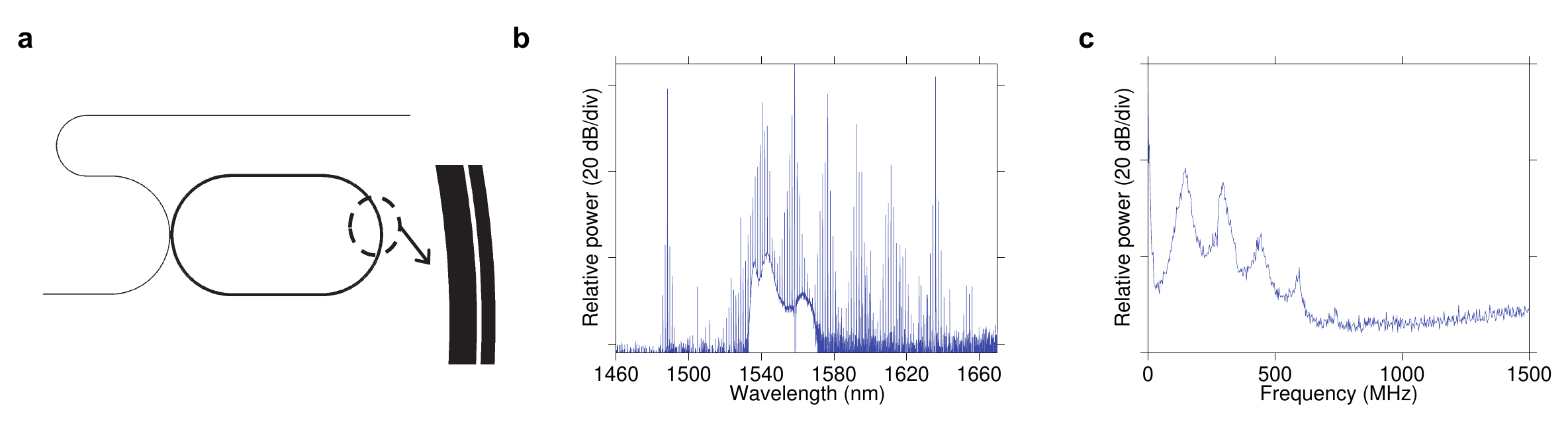}}
\caption{
{\bf Type~II frequency comb with a concentric-racetrack-resonator.}
{\bf a,} Layout image of the concentric-racetrack resonator.
Geometric parameters are $h=300$~nm, $R_{\rm out}=100~\mu$m, $g=500$~nm, $w_{\rm out}=1400$~nm, $w_{\rm in}=2800$~nm, and $L_r=150~\mu$m.
{\bf b,} Type~II comb spectrum and corresponding {\bf c,} RF intensity noise.
}
\label{figS_type2}
\end{center}
\end{figure}
Figure~\ref{figS_type2}a shows the layout image of the concentric-racetrack resonator, whose bending radius and widths are increased to
$R_{\rm out}=100~\mu$m, $w_{\rm out}=1400$~nm, and $w_{\rm in}=2800$~nm, respectively.
Other geometric parameters are $h=300$~nm, $g=500$~nm, and $L_{\rm st}=150~\mu$m.
To increase the coupling efficiency between bus waveguide and resonators, we bent the bus waveguide with the same bending radius of $R=100~\mu$m.
We were able to generate the frequency combs, and Figs~\ref{figS_type2}b and \ref{figS_type2}c show the corresponding comb spectrum and RF noise, respectively.
The generated comb exhibits non-native mode spacing (a type~II comb), with high RF intensity noise attributed to beating between sub-combs \cite{ferdous2011spectral,herr2012universal}.

\section{Line-by-line pulse shaping and autocorrelation measurements}
\begin{figure}[!ht]
\begin{center}
{\includegraphics[width=1.0\textwidth]{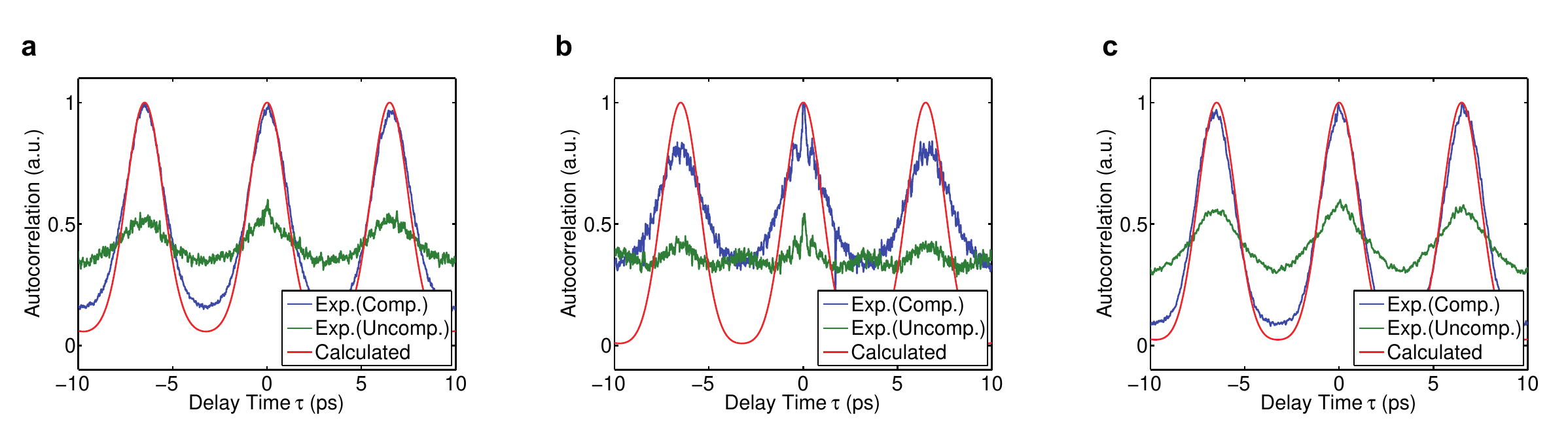}}
\caption{
{\bf Time domain autocorrelation (AC) measurements.}
{\bf a, b, c,}
Time domain AC measurements of the comb state III in Figs.~3e, 3f, and 3g, respectively.
Phase correction via line-by-line pulse shaping is performed.
Autocorrelations with (blue) and without (green) phase correction are displayed.
The autocorrelations calculated based on the experimental power spectrum under the assumption of flat spectral phase are also shown (red).
}
\label{figS_AC}
\end{center}
\end{figure}
To test the coherence of comb states III in Figs.~3e, 3f, and 3g from the main manuscript,
we selected a subset of the comb lines of each comb, conducted line-by-line pulse shaping experiments, and measured the autocorrelation (AC) \cite{ferdous2011spectral};
figures~\ref{figS_AC}a, \ref{figS_AC}b, and \ref{figS_AC}c are the corresponding AC results, respectively.
Green and blue curves in each figure are the experimentally measured AC traces before and after the phase correction, respectively,
and red is the calculated AC that is based on the measured power spectrum and assuming flat spectral phase.
In Figs.~\ref{figS_AC}a and \ref{figS_AC}c, notice the clear difference between before (green) and after (blue) the phase correction;
the phase corrected results (blue) are close to the calculated ACs (red).
These data indicate the coherence of comb states III in Figs.~3e and 3g, as we anticipated from the low RF noise \cite{ferdous2011spectral}.
However, in Fig.~\ref{figS_AC}b, a clear difference between the phase corrected (blue) and calculated (red) ACs remain,
even with the phase correction via line-by-line shaping.
In particular, the experimental AC trace shows very poor contrast ratio
compared both to the calculated curve and to the AC traces in Figs.~\ref{figS_AC}a and \ref{figS_AC}c.
Poor contrast ratio in intensity autocorrelation is a hallmark of broadband intensity noise \cite{ferdous2011spectral,weiner2011ultrafast}.
Thus, the autocorrelation data indicate that the coherence of comb state III in Fig.~3f is poor, consistent with the high RF noise observed for this comb.

\bibliography{CRR_arxiv_SI}